\documentclass[dvips,10pt]{article}
\usepackage{color,graphicx}
\usepackage{setspace}
\usepackage{tabularx}
\newcolumntype{C}{>{\centering}X}
\title{Quantum Logic Processor: A Mach Zehnder Interferometer based Approach}
\author{Angik Sarkar\\Department of Electrical Engineering\\Indian Institute of Technology,Kharagpur,India\\\\Ajay Patwardhan\\
St.Xaviers College,Mumbai,India
\\\\and
\\\\T.K.Bhattacharyya\\
Department of E \& ECE\\Indian Institute of
Technology,Kharagpur,India}
 \date{}
 
\begin{document}

   \maketitle
   \begin{abstract}
 Quantum Logic Processors can be implemented with Mach Zehnder Interferometer(MZI)
 configurations for the Quantum logic operations and gates.
In this paper, its implementation for both optical and electronic
system has been presented. The correspondence between Jones
matrices for photon polarizations and Pauli spin matrices for
electrons gives a representation of all the unitary matrices for
the quantum gate operations. A novel quantum computation system
based on a Electronic Mach Zehnder Interferometer(MZI) has also
been proposed. It uses the electron spin as the primary qubit.
Rashba effect is used to create Unitary transforms on spin qubits.
A mesoscopic Stern Gerlach apparatus can be used for both spin
injection and detection. An intertwined nanowire design is used
for the MZI. The system can implement all single and double qubit
gates. It can easily be coupled to form an array. Thus the Quantum
Logic Processor (QLP) can be built using the system as its
prototype.
\end{abstract}

\maketitle
\newpage
\section{INTRODUCTION}
According to the ITRS (International Technology Roadmap for
Semiconductors) Roadmap~\cite{ITRS}, the channel length of CMOS
should diminish to 22nm by the year 2015 to maintain the ever
increasing computing demands. As dimensions shrink further, the
atomistic limitations will come into light and Boolean Logic will
start to fail. To continue further scaling and improve computing
power, Quantum Logic will become inevitable. Nevertheless the
experimental realization of quantum logic has not been so
impressive.

Preskill~\cite{Preskill} has estimated that to have a reliability
of 10$^{-6}$ atleast 10$^6$ qubits must be present in the quantum
logic system. Such a large number of qubits is easily possible in
a solid state system only. Thus, tremendous improvements have been
made in the field of solid state quantum computation in the last 7
years e.g. Nuclear Magnetic Resonance
(NMR)~\cite{Chuang},electrons floating on liquid
helium~\cite{helium}, quantum dots~\cite{Loss}, terahertz cavity
quantum electrodynamics~\cite{terahertz}, Cooper-pair
box~\cite{Cooper}, superconducting quantum interference
loop~\cite{super}, ion trap quantum computer~\cite{Monroe} spin in
silicon~\cite{Kane} etc. Nevertheless the experimental realization
of solid state systems are still nowhere near Preskill's vision.
In this project we devise a scheme of implementation of quantum
logic in an electronic Mach Zehnder Interferometer.

This paper has been arranged as follows-a fleeting glance of
quantum computing and MZI has be given in the beginning. This will
be followed by the implementation of various quantum logic gates
using the optical MZI. Then we show the transformation of the
scheme used in optical MZI to electronic MZI. Finally, the scheme
of implementation of quantum logic using electronic MZI will be
detailed.

\section{QUANTUM COMPUTING: AN OVERVIEW}

\begin{figure}[h]
\centering
    \includegraphics[bb=90 85 335 290,height=6cm]{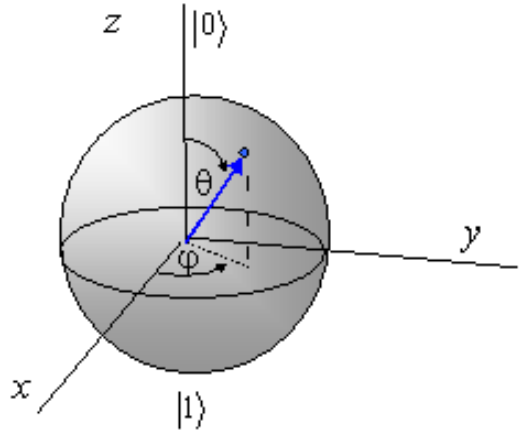}
    \caption{The Bloch Sphere}
    \label{BlochSphere}
\end{figure}

Quantum computation and information is the study of the
information processing tasks that can be accomplished using
quantum mechanical systems. But it may be argued that today's
computers are using nano-sized components where quantum effects
play a big role. So are these computers built on 'quantum
mechanical systems' quantum computers? The answer is NO. It is
because the logic on which the computer operates is classical
rather than quantum mechanical.\\\indent Just as the classical
computation is built upon bits, quantum computation also has an
analogous concept called qubits. The main difference between a bit
and a qubit is that while a bit can be either 0 or 1, a qubit can
be in a superposition of states $\left|0\right\rangle $ and
$\left|1\right\rangle$  (where $\left|~ \right\rangle$ is the
Dirac notation).
                  It is possible to form a linear combination of states called superposition:
\begin{equation}
\left|\Psi\right\rangle = \alpha \left|0\right\rangle + \beta
\left|1\right\rangle
\end{equation}
$\alpha$ and $\beta$ are complex numbers such that
$\left|\alpha\right|^2 + \left|\beta\right|^2 = 1$. A qubit can be
geometrically represented by a Bloch sphere as seen in Fig.
\ref{BlochSphere}.
Analogous to classical computation, the operations on qubits are carried out using quantum logic gates.

\section{MZI: AN OVERVIEW}

\begin{figure}[htbp]
\centering
    \includegraphics[height=6cm]{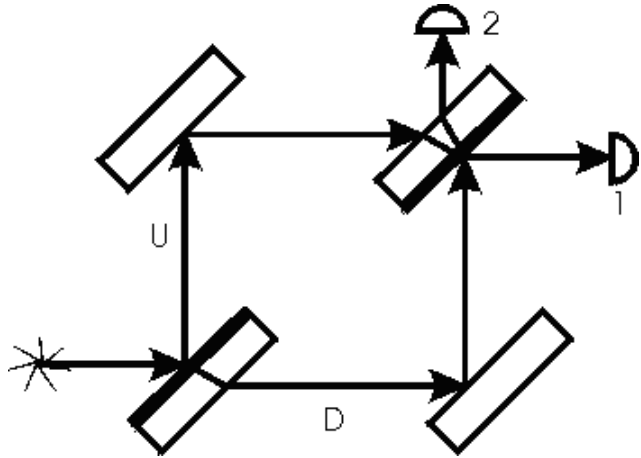}
    \caption{The optical Mach Zehnder Interferometer}
    \label{opticalMZI}
\end{figure}

\begin{figure}[h]
    \centering
        \includegraphics{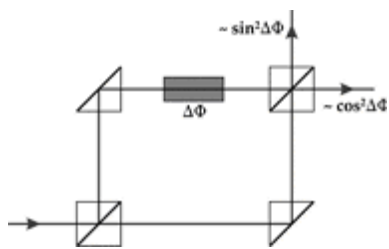}
    \caption{MZI with Phase Shifter}
    \label{fig:MZI+Phaseshift}
\end{figure}

The ubiquitous Mach Zehnder Interferometer was discovered almost a
century ago. The simple structure is as shown in Fig
\ref{opticalMZI} \cite{MZI}.
\newpage
\begin{center}
\begin{tabular}{|l|c|}\hline
  \centering\textbf{ELEMENT} & \textbf{SYMBOLS}\\\hline\hline
  Light Source &

    \includegraphics{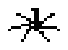}
\\\hline
50-50 Beam Splitter &

    \includegraphics{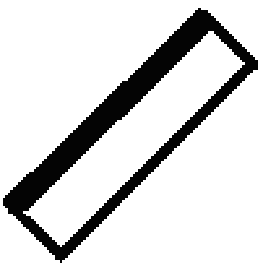}
\\\hline
Totally reflecting mirror &

    \includegraphics{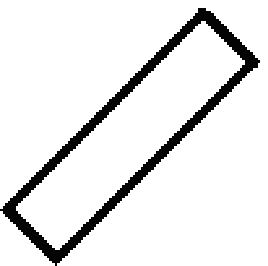}
\\\hline
Detector &

    \includegraphics{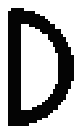}
\\\hline
\end{tabular}
\end{center}
Different elements can be put in either of the paths of Mach
Zehnder Interferometer for manipulating the output states. Fig
\ref{fig:MZI+Phaseshift}. shows a phase shifter in one of the
paths.
The simple structure has led to numerous uses of the optical MZI
since its
inception.\cite{Martin,Winckler,Ludman,MZIex1,MZIex2,Jonathan,Jacke}.
It is this flexibility that makes it ideal for quantum computation
also.

\section{QUANTUM LOGIC IN OPTICAL MZI}

The implementation of quantum logic based on optical MZI harnesses
photon polarization as qubit. The different polarizations are
represented vectorially using Jones vectors\cite{Jones}.The
different Jones vectors are tabulated in Table
\ref{tab:JonesVectors}.

\begin{table}[h]
    \centering
        \begin{tabular}{|l|c|}\hline
            \centering \textbf{POLARIZATION} & \textbf{JONES VECTOR}\\\hline\hline
            Linear horizontal & $\left[\begin{array}{c}1\\0\end{array}\right]$\\\hline
            Linear vertical & $\left[\begin{array}{c}0\\1\end{array}\right]$\\\hline
            Linear at $+45^{o}$ & $\frac{1}{\sqrt{2}}\left[\begin{array}{c}1\\1\end{array}\right]$\\\hline
            Linear  at $-45^{o}$&$\frac{1}{\sqrt{2}}\left[\begin{array}{c}1\\-1\end{array}\right]$\\\hline
            Circular, right-handed &$\frac{1}{\sqrt{2}}\left[\begin{array}{c}1\\-\iota\end{array}\right]$\\\hline
            Circular, left-handed &$\frac{1}{\sqrt{2}}\left[\begin{array}{c}1\\\iota\end{array}\right]$\\\hline
        \end{tabular}
    \caption{Jones Vectors}
    \label{tab:JonesVectors}
\end{table}
The elements that are used for manipulating the polarization
stated are represented mathematically using Jones Matrices. The
Jones Matrices for various optical elements are tabulated in Table
\ref{tab:JonesMatrices}.

\begin{table}[h]
    \centering
        \begin{tabular}{|l|c|}\hline
            \centering \textbf{POLARIZATION} & \textbf{JONES MATRIX}\\\hline\hline
            Linear horizontal polarizer& $\left(\begin{array}{cc}1&0\\0&0\end{array}\right)$\\\hline
            Linear vertical polarizer&$\left(\begin{array}{cc}0&0\\0&1\end{array}\right)$\\\hline
            Linear polarizer at $+45^{o}$&$\frac{1}{2}\left(\begin{array}{cc}1&1\\1&1\end{array}\right)$\\\hline
            linear polarizer at $-45^{o}$ &$\frac{1}{2}\left(\begin{array}{cc}1&-1\\-1&1\end{array}\right)$\\\hline

            Quarter-wave plate,fast axis vertical &$\exp{i\pi/4}\left(\begin{array}{cc}1&0\\0&-i\end{array}\right)$\\\hline

            Quarter-wave plate, fast axis horizontal &$\exp{i\pi/4}\left(\begin{array}{cc}1&0\\0&i\end{array}\right)$\\\hline

            Circular polarizer, right-handed &$\frac{1}{2}\left(\begin{array}{cc}1&i\\-i&1\end{array}\right)$\\\hline

            Circular polarizer,left-handed &$\frac{1}{2}\left(\begin{array}{cc}1&-i\\i&1\end{array}\right)$\\\hline

            Beam Splitter&$\frac{1}{\sqrt{2}}\left(\begin{array}{cc}1&-1\\1&1\end{array}\right)$\\\hline

        \end{tabular}
    \caption{Jones Matrices}
    \label{tab:JonesMatrices}
\end{table}
\clearpage
 The implementation of various quantum logic gates using optical
 MZI is tabulated in Table. \ref{tab:QuantumLogicGates}.
\begin{table}[h]
    \centering
        \begin{tabular}{|l|c|c|l|}\hline
            \textbf{Quantum Logic Gate} & \textbf{Unitary Matrix}   & \textbf{Relation for MZI} &\textbf{Elements}\\&&\textbf{implementation}&\\\hline

            Beam Splitter(B($\theta$)) & $\left[\begin{array}{cc}cos\theta & -sin\theta \\ sin\theta & cos\theta\end{array}\right]$&&\\\hline
            50-50 Beam Splitter(B) & $\frac{1}{\sqrt{2}}\left[\begin{array}{cc}1&-1\\1&1\end{array}\right]$&&\\\hline

            Hadamard(H) & $\frac{1}{\sqrt{2}}\left[\begin{array}{cc}1&1\\1&-1\end{array}\right]$&H=BZ&50-50 Beam splitter\\\hline

            Phase flip gate (Z)& $\left[\begin{array}{cc}1&0\\0&-1\end{array}\right]$&Z=HB& $\pi$ Phase shifter\\\hline

            Bit Flip gate (X) & $\left[\begin{array}{cc}0&1\\1&0\end{array}\right]$ & X=BH & Beam Splitter,\\&&&Hadamard\\\hline

            T gate & $\left[\begin{array}{cc}1&0\\0&\exp(i\pi/4)\end{array}\right]$ && $\pi$/4 phase shifter\\\hline

            S gate & $\left[\begin{array}{cc}1&0\\0&i\end{array}\right]$ && Quarter wave plate\\\hline

            Pauli Y gate & $\left[\begin{array}{cc}0&-i\\i&0\end{array}\right]$&&\\\hline

            CNOT gate & $\left[\begin{array}{cccc}1&0&0&0\\0&1&0&0\\0&0&0&1\\0&0&1&0\end{array}\right]$&(I$\otimes$H)$\times$K$\times$(I$\otimes$H) & $\begin{array}{c}\textnormal{Kerr Media(K)}\\\textnormal{Hadamard(H)}\\\textnormal{Identity(I)}\end{array}$\\\hline

        \end{tabular}
    \caption{Quantum Logic Gates}
    \label{tab:QuantumLogicGates}
\end{table}
\newpage

\section{ANALOGY BETWEEN ELECTRONS AND PHOTONS}

The analogy and similarity between electrons and photons has been
noted in Table \ref{tab:ComparisonBetweenElectronsAndPhotons}.

\begin{table}[h]
    \centering
        \begin{tabular}{|c|c|}\hline
            \textbf{PHOTONS} & \textbf{ELECTRONS}\\\hline\hline
            Electric Field(\textbf{E}) & Wavefunction(\textbf{$\Psi$})\\\hline
            Polarization & Spin\\\hline

            Poynting Vector(\textbf{P})&Current Density(J)\\$\approx$ Re [\textbf{E}$^{*}\times$\textbf{H}]& $\approx$ Re[i$\Psi^{*}\times \nabla\Psi$]\\~ ~ $\approx$ Re [-i\textbf{E}$^{*}\times$\textbf{$\nabla\times$E}]&\\\hline

            exp(-i$\omega$t)& exp(-iEt/$\hbar$)\\
            $\omega$-Frequency& E-Energy\\\hline

            $\nabla^{2}$\textbf{E} = $\omega^{2}\mu\epsilon$\textbf{E}& $\nabla^{2}\Psi$ = -(2m/$\hbar^{2}$[E-U]$\Psi$\\\hline

            k$^{2}$ = $\omega^{2}\mu\epsilon$ & k$^{2}$ = -(2m/$\hbar^{2}$[E-U]\\\hline

        \end{tabular}
    \caption{Comparison between Electrons and Photons}
    \label{tab:ComparisonBetweenElectronsAndPhotons}
\end{table}
The similarity of photon polarization and electron spin on which this whole concept of a spin MZI is based must be illustrated further. The comparison between electron spin and photon polarization is shown in Table \ref{tab:ComparisonBetweenElectronSpinAndPhotonPolarisation}.\\

\begin{table}[h]
    \centering
        \begin{tabular}{|l|l|}\hline

        \textbf{PHOTON POLARIZATION}& \textbf{ELECTRON SPIN}\\\hline\hline
        Spinor & Spinor\\\hline

        Jones vector&   Spin polarization vector\\\hline
Spin $\pm$1 &Spin $\pm$1/2\\\hline
Jones matrices &Pauli matrices\\\hline

        \end{tabular}
    \caption{Comparison between Electron Spin and Photon polarization}
    \label{tab:ComparisonBetweenElectronSpinAndPhotonPolarisation}
\end{table}

\newpage

The realization of various gates and the relation between Jones
and Pauli matrices has been shown in the Table
\ref{tab:JONESPAULIAComparison}.

\begin{table*}[htbp]
    \centering
\begin{tabular}{|c|c|c|c|}\hline
 Optical Element & Jones Matrix & Pauli Matrix & Quantum Logic Gate\\\hline\hline
 Linear horizontal polarizer& $\left(\begin{array}{cc}1&0\\0&0\end{array}\right)$&$I+\sigma_{3}$& \\\hline
Linear vertical polarizer&$\left(\begin{array}{cc}0&0\\0&1\end{array}\right)$& $I-\sigma_{3}$& \\\hline
Linear polarizer at $+45^{o}$&$\frac{1}{2}\left(\begin{array}{cc}1&1\\1&1\end{array}\right)$&$\frac{1}{2}[I+\sigma_{1}]$ & \\\hline
linear polarizer at $-45^{o}$&$\frac{1}{2}\left(\begin{array}{cc}1&-1\\-1&1\end{array}\right)$&$\frac{1}{2}[I-\sigma_{1}]$ & \\\hline
Quarter-wave plate,&&&\\fast axis vertical&$\exp{i\pi/4}\left(\begin{array}{cc}1&0\\0&-i\end{array}\right)$& & \\\hline

Quarter-wave plate,&&&\\ fast axis
horizontal&$\exp{i\pi/4}\left(\begin{array}{cc}1&0\\0&i\end{array}\right)$&
Phase Gate
& \\\hline Circular polarizer,&&&\\
right-handed&$\frac{1}{2}\left(\begin{array}{cc}1&i\\-i&1\end{array}\right)$&
$\frac{1}{2}[I-\sigma_{2}]$& \\\hline Circular polarizer,&&&\\
left-handed&$\frac{1}{2}\left(\begin{array}{cc}1&-i\\i&1\end{array}\right)$&$\frac{1}{2}[I+\sigma_{2}]$
& \\\hline Beam
Splitter&$\frac{1}{\sqrt{2}}\left(\begin{array}{cc}1&1\\1&-1\end{array}\right)$&
$\frac{1}{\sqrt{2}}[\sigma_{1}+\sigma_{3}]$ & Hadamard Gate
\\\hline &&$\sigma_{1}$ & X Gate\\\hline
&&$\sigma_{3}$ & Z Gate\\\hline

\end{tabular}
\caption{Jones and Pauli matrices,a comparison}
\label{tab:JONESPAULIAComparison}
\end{table*}

In Table \ref{tab:JONESPAULIAComparison}. $\sigma_{i}$ where i={1,2,3) are Pauli Matrices.\\
The four Pauli Matrices are:\\
\begin{displaymath}
I=\sigma_{0} = \left(
  \begin{array}{cc}
  1&0\\0&1\end{array}\right)
  \end{displaymath}
\begin{displaymath}
  \sigma_{1} = \left(
  \begin{array}{cc}
  0&1\\1&0\end{array}\right)
  \end{displaymath}
  \begin{displaymath}
  \sigma_{2} = \left(
  \begin{array}{cc}
  0&-i\\i&0\end{array}\right)
  \end{displaymath}
  \begin{displaymath}
  \sigma_{3}= \left(
  \begin{array}{cc}
  1&0\\0&-1\end{array}\right)
  \end{displaymath}

Interestingly, $\sigma_{1}$ is NOT (X) gate and $\sigma_{3}$ is Z gate.

\section{QUANTUM COMPUTATION WITH ELECTRONIC MZI}

Recently, a great interest has been generated in various
electronic analogues of optical instruments viz. -the electronic
double slit interferometer
\cite{Jonsson,LMarton,Marton,Merli,Tonomura,Yacoby}; spin
dependent Fabry Perot Interferometer \cite{Egger,Pescia};
electro-optic modulator \cite{Datta} etc. This is mainly due to
exhibition of quantum phase coherence in electronic interference
experiments by Aharonov-Bohm oscillations, persistent currents
(PC), weak localization, universal conductance fluctuation etc.

This interest has also been extended to MZI. The ubiquitous
optical MZI was rediscovered when simulation of quantum logic with
optical MZI was presented \cite{Cerf}. An electronic analogue of
the MZI was fabricated recently \cite{Ji} using quantum Hall edge
states. In this paper we will extend this interest in MZI with our
proposal of an electronic MZI to implement spin based quantum
logic gates.

The different elements that can be used for the implementation of
quantum logic using a spin MZI is tabulated in Table
\ref{elements}.
\begin{table}[htbp]
\centering
    \begin {tabularx}{\linewidth}{|*{3}{C|}}\hline
        \textbf{Optical Element}&   \textbf{Electronic element}&\textbf{Unitary matrix}\tabularnewline\hline\hline

        Polariser/Analyser & Ferromagnetic material which produces spin sub band splitting. The spin which is to be selectively transmitted should have $\vec{P_{0}}$ parallel to $\vec{M}$&Depends on the angle of polarization. However mostly it is like a linear horizontal polarizer
        $\left[\begin{array}{cc}1&0\\0&0\end{array}\right]$
        \tabularnewline\hline

   Phase Shifter &
   \begin{itemize}
        \item Rashba spin orbit interaction
        \item Aharonov Bohm phase
        \item Ferromagnetic material where $\vec{P_{0}}$ is perpendicular to $\vec{M}$
        \end{itemize} & The general matrix of a phase shifter which shifts the phase by $\theta$ is
given by
$\left[\begin{array}{cc}\exp(-i\theta/2)&0\\0&1\end{array}\right]$
(up to an unimportant global phase factor).

\tabularnewline\hline

Polarizing Beam Splitter (PBS)& An arrangement with two beam
splitters and a phase shifter.\cite{Radu}&
$\left|\sigma;k\right\rangle \longrightarrow
cos\theta\left|\sigma;k\right\rangle +
isin\theta\left|\sigma;1-k\right\rangle$ For a beam splitter where
$\sigma$ gives the spin degree of freedom and k gives the  orbital
degree of freedom.\tabularnewline\hline

Wave guides& Electrons in 2-DEG&\tabularnewline\hline

Beam Splitter &
\begin{itemize}
        \item Point contacts
        \item Simple differentiating lines
\end{itemize} &\tabularnewline\hline

\end{tabularx}

\caption{Elements required for implementation of spin MZI}
\label{elements}
\end{table}

\section{DEVICE CONFIGURATION}
Electron spin has been used as the primary qubit in our proposed
device. All single qubit gates can be realized on the spin qubit
using Rashba interaction. However, using the spatial degree of
freedom in a electronic Mach Zehnder Interferometer (MZI), the two
qubit gates have also been realized. The MZI has been assumed to
be free of spin scattering. To ensure this the MZI is expected to
be formed by two intertwined ballistic nanowires.
\subsection{Design Challenges}
There were quite a many challenges in the design of this system. The challenges are discussed one by one.\\
\indent Magnetic field is normally required for the manipulation
of electron spin. Almost all the spin manipulation based quantum
computation systems propose to  use external magnetic field.
Nonetheless, having a different magnetic field in each of the
units of the quantum computation system is lithographically very
challenging. However,a localized magnetic field in a particular
region can be created with the modern technology. Thus the system
should be designed so that it should be more or less free of the
dependence on the magnetic field. An external magnetic field at a
few specified places may be used.\\\indent Spin-polarized
electrons have been traditionally created in semiconductors simply
by illuminating the material with circularly polarized light.
Nonetheless, a purely electrical method for injecting
spin-polarized electrons into semiconductors is needed to
guarantee the success of quantum computing system. In literature,
two different concepts have been employed to solve the problem.
The first approach involves injecting spins from a dilute magnetic
semiconductor(DMS) that acts as an efficient spin aligner when an
external magnetic field is applied. This concept works well at low
temperatures - almost all the electrically injected electrons have
their spins pointing in the same direction. However, it is
extremely difficult to implement at room temperature because most
of the known magnetic semiconductors lose their spin-aligning
characteristics just above liquid-helium temperatures (i.e. above
4 K). The second approach involves injecting spin-polarized
electrons from a ferromagnetic material~\cite{Datta}, where almost
all of the conducting electrons are intrinsically aligned.
However, this approach also faces problems. Randomly oriented
spins - known as magnetically dead layers - in the ferromagnetic
material close to the semiconductor interface are a barrier to
effective spin injection  Hence, a different spin injection and
detection scheme must be proposed to ensure the robustness of the
quantum computing system.\\\indent CNOT gate is an essential gate
for the flexibility of a quantum computation system. Hence the
scheme must be able to implement double qubit gates like CNOT.

The strategies adopted for overcoming all the aforementioned
challenges are discussed in the following section. A localized
magnetic field was created with the aid of Rashba effect using a
localized electric field created by a gate. The spin injection and
detection problem was solved with the aid of a recently proposed
Stern Gerlach apparatus. CNOT gate was also implemented using spin
as the target qubit and spatial degree of freedom as the control
qubit.

\subsection{Rashba Effect in ballistic MZI}
Even in the absence of a magnetic field, the spin degeneracy may
be lifted due to the coupling of the electron spin and its orbital
motion. This mechanism is popularly referred to as the Rashba
effect~\cite{Rashba}. The spin orbit (Rashba) Hamiltonian is given
by
\begin{equation}
H_{R}
=\frac{\alpha}{\hbar}\vec{y}\cdot\left[\vec{\sigma}\times\vec{p}\right]
\end{equation}
Here y axis has been chosen to be perpendicular to the plane of
motion of the electron(the direction of the electric field),
$\alpha$ is the spin-orbit coupling coefficient, $\vec{\sigma}$
represents the Pauli spin matrices, $\vec{p}$ is the momentum
operator. Typical values of $\alpha$ range from $9\times10^{-12}$
eV m at electron density of $n=7\times10^{11} cm^{-2}$ to
$6\times10^{-12} eV m$ at electron density of $n=2\times10^{12}
cm^{-2}$. Due to Rashba interaction, the Fermi sphere splits into
two(see Fig \ref{fig:Fermi_sphere}). Thus a spin dependent band
splitting is achieved.
\begin{figure}[h]
   \centering
        \includegraphics{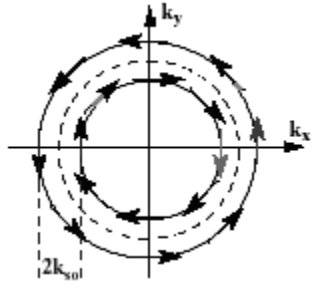}
    \caption{Splitting of the Fermi Sphere}
    \label{fig:Fermi_sphere}
\end{figure}
In the limit of $\alpha=0$, the eigen energies are given by
\begin{equation}
E^{0}_{n}=E_{n}+\frac{\hbar^{2}k_{y}^{2}}{2m^{*}}
\end{equation}
If we treat the spin orbit interaction using the perturbation
model( This model is quite correct as the spin orbit effect is
quite weak), the eigenvalues can be written as:
\begin{eqnarray}
E^{\pm}(k_{y})&=&E^{0}_{n}\pm \alpha k_{y}\\
E^{\pm}(k_{y})&=&E_{n}+\frac{\hbar^{2}k_{y}^{2}}{2m^{*}}\pm \alpha
k_{y}
\end{eqnarray}
Here n is the subband index, $m^{*}$ is the effective mass. The
above equation allows more than one values of $k_{y}$ to have the
same energy. Let these values be $k_{y1}$ and $k_{y2}$. Hence,
\begin{eqnarray}
E^{+}(k_{y1})-E^{-}(k_{y2})&=&\frac{\hbar^{2}}{2m^{*}}(k_{y1}^{2}-k_{y2}^{2})+\alpha(k_{y1}+k_{y2})\\
&=&0\\
k_{y1}-k_{y2}&=&\frac{2m^{*}\alpha}{\hbar^{2}}=\Delta k_{y}\\
\phi_R&=&\Delta k_{y} L\\
\phi_R&=&\frac{2m^{*}\alpha L}{\hbar^{2}}
\end{eqnarray}
Here $\phi_R$ represents the phase shift in the Rashba region.L is
the length of the Rashba region. The unitary transform associated
with this phase shift is
\begin{equation}
U_R=exp[\iota \phi_R \sigma_z]
\end{equation}
Thus the effect of the Rashba interaction is to rotate the spin
direction by $\phi_R$ in the spin space. A point to note here is
that in the above derivation, the intersubband coupling was
neglected. This approximation is valid if the following condition
holds~\cite{Datta}:
\begin{equation}
w\ll\frac{\hbar^2}{\alpha m^*}
\end{equation}
Here w is the width of the ballistic nanowire.

\subsection{Mesoscopic Stern Gerlach Apparatus} As mentioned
earlier, traditionally, ferromagnetic contacts have been used for
spin injection and detection. However it is well known that the
same can be done with a Stern Gerlach apparatus in macroscopic
domain. Extending this concept, spin injection and detection
through a mesoscopic Stern Gerlach Apparatus~\cite{Radu} was
proposed for the quantum computing system.

The aforementioned Stern Gerlach Apparatus uses a MZI like
structure to produce a Polarizing Beam Splitter(PBS). It is a two
input, two output structure and hence is easily compatible with
our MZI design. It has Rashba interaction selectively on one of
the arms. Also there is flux $\Phi$ threading the apparatus. The
outputs can be tuned to have only spin ups and spin downs along
different paths. Let us define the two modes to be k=0,1. Let the
Rashba interaction be present in only the `1' mode. Hence the
transformation can be written as:
\begin{eqnarray}
\left|\uparrow, k \right\rangle &\rightarrow & e^{\iota k \phi_R} \left|\uparrow, k \right\rangle\\
\left|\downarrow, k \right\rangle &\rightarrow & e^{-\iota k
\phi_R} \left|\downarrow, k \right\rangle
\end{eqnarray}
The magnetic flux $\Phi$ threading the interferometer generates a
Aharonov Bohm(AB) phase. This induces a phase difference in the
electronic wavefunctions in the two arms. This phase difference
can be assumed in the `1' mode without any loss of generality.
Thus,
\begin{eqnarray}
\left|\sigma, 0\right\rangle & \rightarrow&\left|\sigma, 0\right\rangle\\
\left|\sigma, 1\right\rangle & \rightarrow& e^{\iota \Phi_{AB}}\left|\sigma, 1\right\rangle\\
\Phi_{AB}&=&\Phi/\Phi_0\\
\Phi_0&=&\frac{hc}{e}
\end{eqnarray}
The net transformation induced by the two mechanisms are as
follows (assuming the beam splitter to be 50-50):
\begin{eqnarray}
\left|\uparrow; 0\right\rangle &\rightarrow & t_0^{\uparrow}\left|\uparrow; 0\right\rangle+t_1^{\uparrow}\left|\uparrow; 1\right\rangle\\
\left|\downarrow; 0\right\rangle &\rightarrow &t_0^{\downarrow}\left|\uparrow; 0\right\rangle+t_1^{\downarrow}\left|\uparrow; 1\right\rangle\\
t_0^{\uparrow,\downarrow}&=&- e^{\iota(\Phi_{AB}\pm\Phi_R)/2} \iota sin\left(\frac{\Phi_{AB}\pm\Phi_R}{2}\right)\\
t_1^{\uparrow,\downarrow}&=& e^{\iota(\Phi_{AB}\pm\Phi_R)/2} \iota
cos\left(\frac{\Phi_{AB}\pm\Phi_R}{2}\right)
\end{eqnarray}
Choosing $\Phi_{AB} = \Phi_R = \pi/2$ ,we get,
\begin{eqnarray}
\left|\uparrow, 0\right\rangle & \rightarrow&\left|\uparrow, 0\right\rangle\\
\left|\downarrow, 1\right\rangle & \rightarrow&
\iota\left|\downarrow, 1\right\rangle
\end{eqnarray}
Hence spin injection can be done as required by choosing an appropriate $\Phi_{AB}$ and $\Phi_{R}$.
 The spin injection can be as high as 100\% as both the spin injector and our device can be of the same material.
 There is no impedance mismatch at the interface.

Single spin detection is a big problem in spintronics. The success
of spin systems depends on efficient spin injection and detection.
However efficient methods for detection of electron spin directly
in solids are still eluding. Nevertheless single charge detection
is possible. Hence some proposals for single spin detection in
nano devices are based on the swap operation of a spin state to a
charge state~\cite{Kane}. Proposals have also been made for spin
detection using Scanning Tunneling Microscopy(STM)~\cite{Manassen}
and Magnetic Resonant Force Microscopy(MRFM)~\cite{Sidles,Rugar}.

Single Electron Transistor(SET) can used to detect single electron
charge. Thus in our system too, efficient single spin detection
can be done by performing the swap operation of a spin state to a
charge state. The MSGA performs the swap function efficiently.
Hence for spin readout in our system, an MSGA is coupled to the
output of the electronic MZI. The output of the MSGA is fed to
SET. Again the MSGA can be fabricated in the same material as MZI.
Hence the spin detection efficiency would be high.

\subsection{Quantum Logic Gates} The realization of different
quantum logic gates will be discussed in this section. The quantum
logic gates can be implemented in an electronic MZI system.A
schematic MZI system with the spatial(modal) and spin qubit has
been shown in Fig \ref{wireMZI}.
\begin{figure}
\centering
\includegraphics[height=6cm]{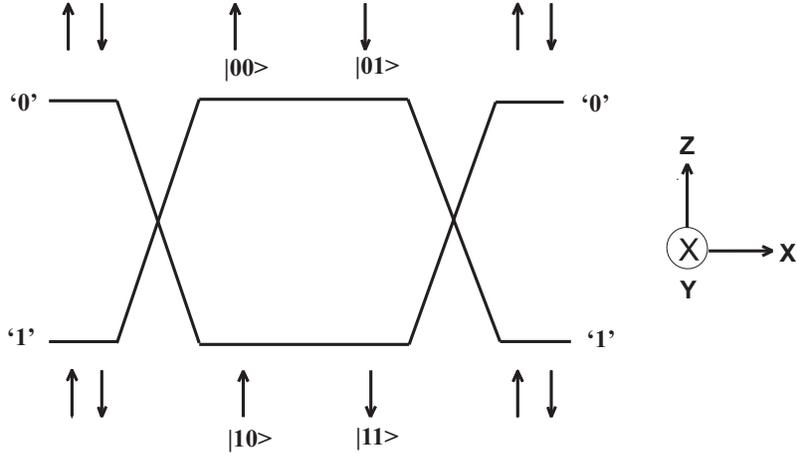}
\caption{A schematic electronic MZI with the different qubits
depicted} \label{wireMZI}
\end{figure}

The electron spin is the primary qubit in the proposed system.
Hence all the single qubit gates have been designed for the spin
qubit. It was mentioned in section 7.2 on Rashba effect, that it
rotates electron spin in the spin space. Now it is well known that
all single qubit gates are nothing but rotations on the Bloch
sphere. An electron spin can easily be visualized on the Bloch
sphere with the $\uparrow$ and $\downarrow$ on opposite poles. All
superposition states can also be represented by points on the
Bloch sphere. Thus rotations in spin space correspond to rotations
on Bloch sphere (Fig \ref{BlochSphere}). It has already been shown
that all rotations can be achieved by varying the spin orbit
coupling coefficient $\alpha$. $\alpha$ in turn can be tuned by a
gate voltage in a ballistic MZI ( Fig \ref{nanoMZI}).
\begin{figure}
\centering
\includegraphics[bb=145 88 510 410,height=8cm]{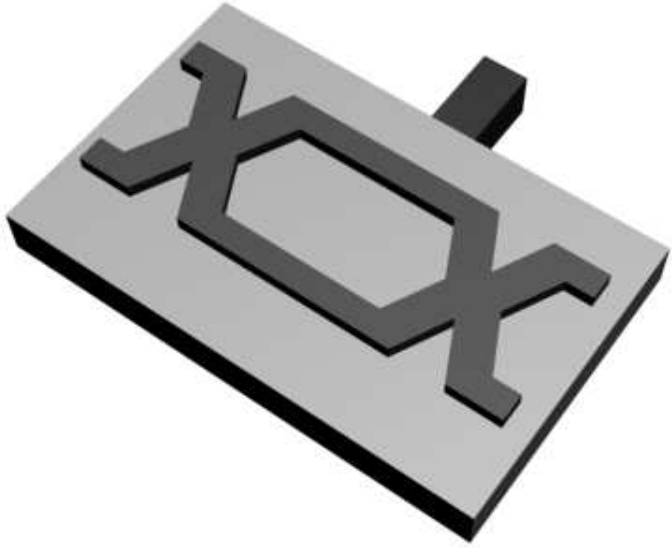}
\caption{A schematic ballistic MZI with gates for application of
voltage. The Rashba interaction can be controlled by tuning the
gate voltages} \label{nanoMZI}
\end{figure}

For an analytical calculation, the device size can be fixed at 50
nm by 30 nm (50 nm is the distance between two consecutive beam
splitters, 30 nm is the distance between the centers of the two
parallel paths.) The width of the channel is assumed to be 5nm,
the length of Rashba interaction is 20nm. In InGaAs/InAlAs system,
a similar structure has been fabricated~\cite{Nitta}. The
effective mass($m^*$) is 0.05 $m_0$($m_0$ is the rest mass of a
free electron) in that system. Also the confinement voltage was
0.53 V . In this system the various $\alpha$ values required for
various gates are shown in Table \ref{alpha}.
\begin{table}
    \centering
        \begin{tabular}{|l|c|c|c|}\hline
        Quantum Logic Gate & Unitary Matrix &$\alpha (\times 10^{-11} eV m)$ \\\hline\hline
    50-50 Beam Splitter(B) & $\frac{1}{\sqrt{2}}\left[\begin{array}{cc}1&-1\\1&1\end{array}\right]$ &1.198\\\hline
    Hadamard Gate(H) & $\frac{1}{\sqrt{2}}\left[\begin{array}{cc}1&1\\1&-1\end{array}\right]$&1.198\\\hline
    Not Gate(X) &$\left[\begin{array}{cc}0&1\\1&0\end{array}\right]$&2.397\\\hline
    Phase flip gate (Z)& $\left[\begin{array}{cc}1&0\\0&-1\end{array}\right]$&4.795\\\hline
    T gate & $\left[\begin{array}{cc}1&0\\0&\exp(i\pi/4)\end{array}\right]$ &0.300\\\hline

            S gate & $\left[\begin{array}{cc}1&0\\0&i\end{array}\right]$&0.599\\\hline
            \end{tabular}
    \caption{The spin orbit coupling coefficient $\alpha$ required for some standard single qubit quantum logic gates}
    \label{alpha}
\end{table}

The values are close to the values obtained in experiments
~\cite{Nitta,Grundler}. Hence the single qubit gates can be easily
obtained by the rotation of the spin.

Double qubit gates are an imperative feature of any quantum
computing system. To achieve double qubit gates, entanglement has
to be obtained. The electrons are constrained to move in either of
the two modes, `0' or `1'. Hence the spatial location of the
electron(the modes) can also used as a qubit. The beam splitter
acts on the spatial qubit and can be represented by the U(2)
matrix,
\begin{equation}
\left|k;\sigma\right\rangle \rightarrow \cos \theta
\left|k;\sigma\right\rangle + \iota \sin\theta
\left|1-k;\sigma\right\rangle
\end{equation}

Hence the superposition states for the spatial qubit can also be
obtained. As shown in Fig \ref{wireMZI}, the two qubit
representations can be depicted with the spatial qubit as the 1st
qubit and spin as the next qubit. The first qubit can easily act
as the control qubit. If the Rashba field is turned on in the `1'
mode only, spin manipulation will occur in that mode only. For
example, in the CNOT, gate, the Rashba field is turned on in mode
`1' only with $\alpha $ as shown in Table \ref{alpha} for NOT
gate. Thus there is preferential flipping of the spin qubit in the
`1' mode only. This is simply the CNOT gate. Similarly controlled
Z gate etc can also be obtained. The Bell states can easily be
obtained in this system. This will require a Hadamard on the
spatial qubit and then a CNOT gate with the spatial qubit as the
control qubit.

\section{Conclusions}
In this paper, Mach Zehnder based quantum computing systems have
been presented. All single qubit and double qubit quantum logic
gates are feasible in this system. The system can be implemented
in a ballistic nanowire MZI system. The specifications suggest
that it can be made within the current state-of-the-art
technological facilities available.

The device size should be smaller than the phase coherence and
spin coherence length. These have been typically reported to be 20
microns~\cite{phase} and 100 microns~\cite{spin} respectively.
Hence an array of about 1000 MZI units can be placed in an array
to perform any complex operation. This could lead to the ultimate
Quantum Logic Processor (QLP). The problems of coupling that exist
in a static electron spin quantum computation system have been
eliminated in our proposed mobile spin qubit representation. Hence
it is better suited for implementation of the QLP.

\section{Acknowledgements}
Angik Sarkar and Ajay Patwardhan acknowledge the help of the NIUS
Programme, HBCSE, Mumbai,India.


\end{document}